\documentstyle[prl,multicol,psfig,aps]{revtex}
\begin{document}
\draft

\title{Evolution of a polaron band through the phase diagram of 
Nd$_{2-x}$Ce$_x$CuO$_{4-y}$}
\author{S. Lupi, P. Maselli, M. Capizzi, and P. Calvani}
\address{Istituto Nazionale di Fisica della Materia - Dipartimento di Fisica, 
Universit\`a di Roma ``La Sapienza'', Piazzale A. Moro 2, I-00185 Roma, Italy}
\author{P. Giura and P. Roy}
\address{Laboratoire pour l'Utilization de Rayonnement Electromagn\'etique,
Universit\'e Paris-Sud, 91405 Orsay, France}
\date{\today}
\maketitle

\begin{abstract}
The evolution with doping and temperature of the polaronic absorption in a
family of high-T$_c$ superconductors is first followed here across the whole 
phase
diagram. The polaron band softens with increasing doping, is still present in 
the superconducting phase, and vanishes in the overdoped metallic phase.
These results point toward the existence of polaron aggregates at low
temperature and provide an experimental ground for the increasing
number of theoretical descriptions of many-polaron systems.
\end{abstract} 

\begin{multicols}{2}
\narrowtext

An absorption band (often called $d$ band) appears at $\approx$ 1000 cm$^{-1}$ 
in the infrared spectra of many parent compounds of high critical temperature
superconductors (HCTS) upon slight doping by either 
electrons or holes.\cite{Kim,Thomas,NoiPR92,Kastner,Salje} 
The polaronic nature of the $d$ band has been established on sound 
arguments,\cite{Kim,Kastner,Salje,Bucher} which include the observation that 
this band is made
of overtones of infrared active vibrational modes (IRAV's) induced by
doping.\cite{Noissc} The polarons are 
most likely in the Cu-O planes that are common to all cuprates where the $d$
band has been observed. 

Recently, an increasing theoretical effort has been
devoted to the investigation of many-polaron
systems.\cite{Devreese,Cataudella,Quemerais,Lorenzana,Grilli,Yonemitsu}
Some role of polarons in 
high-$T_c$ superconductivity has
also been proposed by several authors.\cite{Emin,Ranninger} 
An experimental study of the 
$d$  band through the whole phase diagram of a cuprate
would provide a useful basis to the above theoretical
efforts. 

In the present paper the above task is 
pursued by following the evolution of the polaron band in ten 
Nd$_{2-x}$Ce$_x$CuO$_{4-y}$ (NCCO) single 
crystals. Their reflectivity 
$R(\omega,T$) has been measured from 10 to 300 K and from 50 to 15000 cm$^{-1}$.
Those measurements allow us to follow with unprecedented detail the 
evolution of the $d$ band across the insulator-to-superconducting transition 
(IST) in a family of cuprates. This procedure also enables one to overcome 
the intrinsic difficulties in discriminating the $d$ band from the Drude 
term in the infrared spectrum of a single, metallic sample.  

This investigation in the electron-doped NCCO should be relevant 
to HCTS, independently of the type of carriers involved. In fact, the well 
known equivalence between the optical properties of electron- and hole-doped 
HCTS (see, e.g., Refs. \onlinecite{Uchida} and \onlinecite{Cooper}) has been 
recently extended to the
anomalous transport properties.\cite{nota,Fournier,Boebinger} 

The main results of the present work are summarized as follows: 
i) the average polaron energy softens 
with increasing doping and/or lowering temperature through the whole insulating 
phase, with an abrupt change of slope an intermediate doping; ii) this behavior 
persists in the metallic superconducting phase, where 
a polaron contribution peaked at a finite energy is still distinguishable, even 
at optimum doping; iii) no change in the polaron energy with doping is detected 
at IST; iv) a plain Drude behavior is observed in the overdoped, not 
superconducting regime. These results are discussed in connection with recent 
models of polaron-polaron 
interactions.\cite{Devreese,Cataudella,Quemerais,Lorenzana,Grilli,Yonemitsu} 
 
The NCCO single crystals investigated here were prepared as described in 
Ref. \onlinecite{NoiPR92}. 
Their doping concentrations are specified in Table I, together 
with that of a sample measured in Ref. \onlinecite{Homes}. The Ce concentration 
was determined as the average of chemical microanalysis measurements at 4 to 7 
positions on the crystal surface. The experimental apparatus has been 
described in 
detail in Ref. \onlinecite{NoiPR98}. The real part of the optical conductivity 
$\sigma(\omega,T)$ has been obtained from Kramers-Kronig transformations of 
$R(\omega,T$)  
measured with the radiation electric field polarized in the Cu-O plane.  
$\sigma(\omega,T)$ is independent of the low- and high-energy extrapolations 
used to extract it,\cite{NoiPR98} at least for $\omega \leq$ 5 $\times$ 10$^3$ 
cm$^{-1}$. 

$R(\omega,T)$, reported in Fig. 1, illustrates how the reflectivity evolves 
with doping for the most representative samples. The peak above 10$^4$ 
cm$^{-1}$ is due to the well known charge transfer (CT) band. A strong 
temperature dependence is observed in the $R(\omega,T)$ of the insulating
samples $C$ ($x=0; y<0.01$), $F$ ($x=0.04; y\sim0$), and $H$ ($x=0.12; y\sim0$)
for $\omega \alt$ 6 $\times$ 10$^3$ cm$^{-1}$. This holds too for
the superconducting sample $I$ ($x$=0.17), while no temperature dependence 
is observed in sample $J$ 
($x$=0.21). The reflectivity of the other samples in Table I smoothly 
interpolates that of the samples shown in Fig. 1. 

\begin{figure}
{\hbox{\psfig{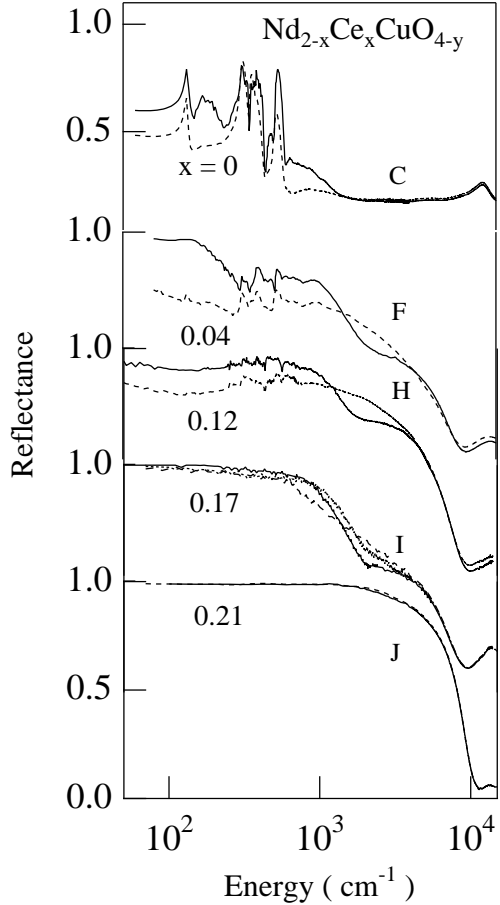}}}
\caption{$R(\omega,T)$ spectra taken at 300 (dashed lines) and 20 K 
(full lines) but for the $x$ = 0.17 sample, which has been measured at low T 
both in the 
superconducting (T=10 K, full line) and in the metallic anomalous phase 
(T=40 K, dotted line).}
\label{1}
\end{figure}

The dependence of the polaron band on T and doping can be
isolated from the plain effect of an increased carrier density by looking at
the renormalized conductivity $\sigma^*(\omega,T)$ = $\sigma(\omega,T)/n^*$,
where                                      

\begin{equation}
n^* = {2m V\over \pi e^2} \int _0^{\omega^*} \sigma (\omega)\, d \omega
\label{neff}
\end{equation}
  
\noindent
is the spectral weight in the CT gap and is proportional to the number of 
carriers per unit cell. $m$ is the carrier effective mass, assumed here equal 
to the free electron mass $m_0$, V is the volume of the unit cell, and 
$\omega^*$=10000 cm$^{-1}$, an approximate value for the CT energy gap. 
$\sigma^*(\omega,T)$ is reported in Fig. 2 in a reduced infrared range for the 
same samples shown in Fig. 1. 

For fixed T and increasing doping, the spectral weight shifts toward low energy
and narrows until a Drude contribution appears above IST. The phonon peaks, 
increasingly screened, nearly vanish in sample $H$. 

\begin{figure}
{\hbox{\psfig{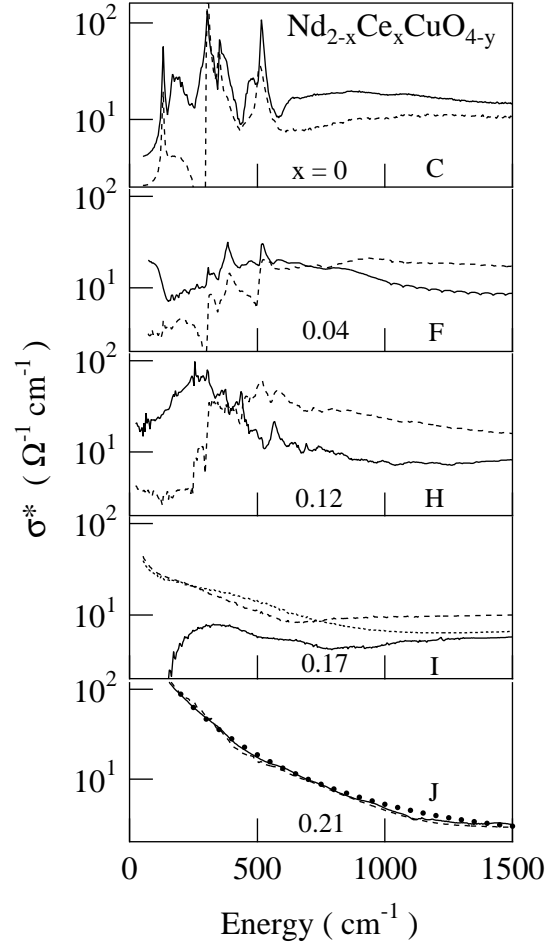}}}
\caption{The optical conductivity per carrier in the unit cell 
$\sigma^*(\omega,T)$ is reported 
in a reduced infrared range for the same samples shown in Fig. 1. Curves at 300
K are given by dashed lines, at 20 K by full lines. For the $x$ = 0.17 sample
$I$, the dotted line refers to 40 K, the full line to 10 K. Finally, for the
$x$ = 0.21 sample $J$, the larger dots represent the fit to a Drude
conductivity.}
\label{2}
\end{figure}

For fixed doping, the spectral weight shifts toward low energy with decreasing 
T until a saturation temperature T$_s$ is reached. Below T$_s$, the far-infrared
optical
conductivity does not change sizably. An inspection to the sets of data 
taken for some samples at intermediate T's suggests that T$_s$ decreases 
with doping. It is equal or greater than room temperature in sample $A$ 
($x$=0; $y<0.005$), on the order of 200 K in sample $D$ ($x$=0; $y<0.04$), 
quite lower in sample $I$ ($x$=0.17; $y<0$) where $\sigma^*(\omega,T)$ still 
changes on going from 40 to 10 K, and finally is not relevant for sample $J$ 
($x$=0.21) 
where no appreciable dependence on T can be detected between 300 and 20 K. A 
finer tuning of the temperature sampling is required in order to get precise 
values of T$_s$ and will be the object of further investigations. Since the 
spectral weight 
undergoes a ``red shift'' for decreasing T, the gap appearing at 300 
K below $\sim 250$ cm$^{-1}$ in sample $H$ is almost completely filled at 20 K,
indicating a low temperature ``metallization'' near IST. 
This ``red shift'' leads also to a dependence on T of the phonon line shapes 
(see the two low-doped samples $C$ and $F$), which can be related 
to a Fano interaction with a polaron continuum shifting from high to low 
energy with decreasing T.\cite{NoiPR98} 

Finally, for the most doped sample $J$, $\sigma^*(\omega,T)$ in Fig. 2
follows a conventional Drude behavior

\begin{eqnarray}
n^*\sigma^* = {\Gamma_D\omega_p^2 \over {\omega^2 + \Gamma_D^2}} \, ,
\label{diel}    
\end{eqnarray}

\noindent
where $\omega_p=(ne^2/m)^{1/2}$ is the plasma frequency and $\Gamma_D$ is 
the inverse of the scattering rate, {\it independent of $\omega$}. In
the same Fig. 2, a best fit of Eq. 2 to data is reported by dots 
($\omega_p$ = 17500 cm$^{-1}$, $\Gamma_D$ = 90 cm$^{-1}$). On the other hand, 
the same type of fit is unsuccessful in sample $I$ where one should add the 
$d$ band to the Drude term in order to satisfactorily fit $\sigma^*(\omega,T)$.

Let us now proceed to a quantitative analysis of $\sigma^*(\omega,T)$. The 
shift of the spectral weight between the insulating samples $C$ and $F$ is 
due to the ``red shift'' of the polaron $d$ band.\cite{NoiPR98} This analysis 
can be extended to more doped samples, where the peak energy of the $d$ band  
increasingly merges with the Drude term, by evaluating the first moment, 
$<\omega>$, of the polaronic contribution to $\sigma(\omega,T)$. In those 
highly doped samples the polaron band can be hardly 
isolated from the Drude and IRAV contributions, while both the mid-infrared 
(MIR)\cite{noiPR92} and CT terms 
can easily be identified. Therefore, we actually estimate $<\omega(T)>$ by 
subtracting from $\sigma(\omega,T)$ the corresponding contributions 
$\sigma_{MIR}(\omega,T)$ and $\sigma_{CT}(\omega,T)$, which
are determined from a standard fit to the $\sigma(\omega,T)$ 
data.\cite{NoiPR98,NoiPR96} In conclusion, an approximate expression for
the first moment of the polaron conductivity is given by 

\begin{equation}
<\omega> \simeq {{\int _0^{\omega^*} \omega\, 
[\sigma -\sigma_{MIR} -\sigma_{CT}]\, d \omega} 
\over
                   {\int _0^{\omega^*} 
[\sigma -\sigma_{MIR} -\sigma_{CT}]\, d \omega}} 
\, .
\label{moment}
\end{equation}

\noindent
This quantity, estimated at T=300 and 20 K, is plotted in 
Fig. 3 vs the spectral weight $n^*$ renormalized to $n_0$, the spectral weight 
in the CT gap of the less doped sample $A$. $<\omega (T)>$ has been estimated 
also for a sample with $x$=0.15 on the grounds of conductivity data at 
10 K from Ref. \onlinecite{Homes} 
and is reported in Fig. 3. 

A major softening of the $d$ band
on going from room to low temperature is confirmed in all samples.
As a function of doping, two different regimes can be identified.
In the diluted polaron regime (samples $A$ to $D$), where the intensity of the 
$d$ band $I_d$ increases\cite{NoiPR96} for $T \to 0$,
$<\omega (T)>$ decreases at a much faster rate than at high doping. In the
latter region, the determination of $I_d$ is affected by large errors. However,
in a few
samples\cite{NoiPR98} $I_d$ seems even to increase with $T$, as predicted 
in Ref. \onlinecite{Kabanov} for a system including polarons, free charges and
impurities.

Quite surprisingly, no abrupt change in $<\omega (T)>$ is observed at 
IST and $<\omega (T)>$ is still finite in the superconducting phase, as
confirmed from the value extrapolated from Ref. \onlinecite{Homes}
(full triangle). $<\omega>$  
vanishes, instead, at the superconducting-to-metal transition SMT, both at 300 
and 20 K, in agreement with the plain Drude behavior of $\sigma(\omega,T)$ in 
sample $J$. It should also be mentioned that the polaron band narrows
continuously for increasing doping, as observed when comparing sample $F$
with sample $H$ in Fig. 2.

\begin{figure}
{\hbox{\psfig{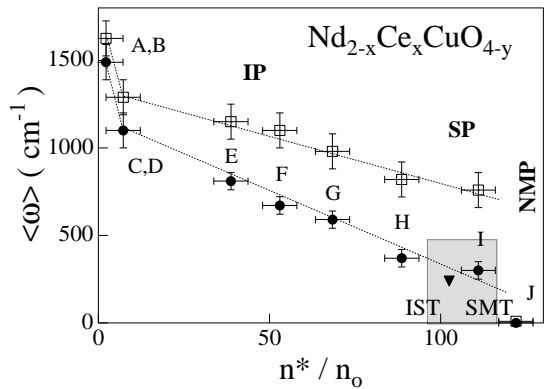}}}
\caption{First moment $<\omega (T)>$ of the polaron $\sigma(\omega)$ at 300 K 
(open squares) and at 20 K
(full dots) for all samples but sample $I$ (measured at 10 K) vs normalized 
values of the effective carrier concentrations $n^*/n^*_0$. A full triangle
gives the value of $<\omega (T)>$ as estimated at 10 K for a sample measured 
in Ref. 21. 
The superconducting phase, SP, is shown by a shaded area. IST
roughly indicates the transition from the insulating phase, IP, to the 
superconducting one. SMT appromaximately indicates the transition from the 
superconducting phase to the normal metal phase, NMP.}
\label{3}
\end{figure}

The above results provide a verification for theoretical models recently
proposed for describing a system of interacting polarons. Indeed, 
$<\omega>$ in Fig. 3 is closely related to the polaron energy 
$E_p$. In the limit of infinite polaron dilution ($n \rightarrow 0$) 
one observes\cite{NoiPR96} an almost symmetric polaron band and 
$<\omega> \simeq$ 2$E_p/\hbar$. At finite doping, both the
polaron size and line shape are unknown and $<\omega>$ remains the best
experimental estimate of $E_p$ presently available.

According to theoretical models accounting for polaron-polaron 
interaction,\cite{Devreese,Cataudella,Quemerais} an increase in the polaron 
density should reduce the electron-phonon coupling. This leads to a decrease of 
$E_p$, in qualitative agreement with present results. 
An explicit expression for the decrease of $E_p$ due to the dipole-dipole 
interaction in a liquid of large Feynman polarons has been recently 
derived:\cite{Lorenzana}  

\begin{equation}
E_p^2 = E_{p0}^2 - 0.6 (q/e) [1+m_0/(m^*_{pol}-m_0)]\Omega_{pol}^2 \, .
\label{lorenzana}
\end{equation}

\noindent
Here, $E_{p0}$ is the polaron energy at infinite dilution,
$(q/e) \sim (1-1/\epsilon_0)$, with $\epsilon_0$ the static dielectric 
constant, $e$ is the free electron charge, $m^*_{pol}$ the polaron effective 
mass, and $\Omega_{pol}^2$ is equal to eight times the integral at the 
denominator of Eq. 3, namely, it is proportional to the polaron density.
Equation (4) quantitatively accounts for the
decrease of the polaron energy from sample $A$ to sample $D$ and predicts that
$E_p$ vanishes at a doping just slightly higher than that of sample $D$,
namely where the rate in the decrease of $<\omega>$ changes; see Fig. 
3. This change of slope implies the insurgence of a new process, e.g., the 
formation of a polaron aggregate, not predicted by any of the above 
theoretical models. This
process should account also for the persistence of polarons in the metallic 
superconducting phase with a collapse of $<\omega (T)>$ in the normal metal.
The narrowing of the polaron band further
suppports the formation of polaron aggregates. These latters may be either
polaron pairs, or clusters, or stripes. 

The implications of a 
hypothetical superconducting polaron-pair aggregate have been discussed and 
compared to the properties of HCTS in a recent study of the tunneling dynamics 
of polarons in a two dimensional antiferromagnet.\cite{Yonemitsu} This model
also predicts a crossover (either sharp or continuous) from the polaron 
aggregate to a Fermi liquid for doping higher than the optimal one.
On the other hand, evidence of stripes in certain HCTS has been provided by a
series of experiments, in particular neutron scattering\cite{Tranquada} and
EXAFS.\cite{Bianconi} Moreover, it can be shown that the  total energy 
of the system decreases if polarons dynamically self-aggregate in 
unidimensional arrays in the Cu-O plane, hereafter called wires, whose spacing 
decreases with 
increasing polaron density.\cite{Grilli} In this context,
at the critical density IST, the Cu-O plane becomes unstable and strong
fluctuations in the wire density together with the increasing interaction 
between wires give rise at low T to a crossover from an insulating phase to 
a superconducting polaron liquid. Therein, localized energy states, monitored
by the survival of the $d$ band, see Fig. 3, are in dynamical equilibrium with 
delocalized free carrier (Drude) states. 
In the superconducting phase, the density of free
carriers should increase superlinearly with doping, since an increase in carrier
density leads to an increase in the screening length with an ensueing further
increase in the number of carriers leaving the wires. This would account for
the abrupt drop of the polaron energy at SMT in Fig. 3, which is 
reached when the screening unbounds all carriers, as 
also suggested in Ref. \onlinecite{Yonemitsu}. 

In summary we have shown that a polaron band dramatically softens in NCCO
for increasing doping and/or decreasing temperature. This band can still be
distinguished from the Drude term in the superconducting phase and disappears
only when $T_c$=0. The well established similarity of both the optical and
transport properties between electron- and hole-doped 
HCTS extends the validity of 
the present results, which then stress the role of polaronic effects in high 
$T_c$ superconductivity.
                                             
We thank W. Sadowski for providing the samples here investigated,  
A. Paolone and B. Ruzicka for collaborating in different phases of 
the data collection, M. Grilli and J. Lorenzana for letting
us know their results prior to publication. We also                       
acknowledge useful discussions with C. Castellani, V. Cataudella, S. Ciuchi, 
J. Devreese, C. Di Castro, S. Fratini, G. Iadonisi, and P. Quemerais.

\end{multicols}

\widetext


\begin{table} 

\caption
{Values of Ce concentration, $x$, and of oxygen nonstoichiometry, $y$, for the 
Nd$_{2-x}$Ce$_x$CuO$_{4-y}$ samples investigated in the present work.}   
\label{Table I}

\begin{tabular}{cccccccccccc}

    &   A    &   B    &   C   &   D   &  E &   F   &   G   &   H   & I$^a$ &   
J      & Homes$^b$\\
\tableline
$x$ &   0    &   0    &   0   &   0   &0.04& 0.04  & 0.10  & 0.12  & 0.17  &
 0.21  &   0.15   \\
$y$ &$<$0.005&$<$0.005&$<$0.01&$<$0.04&$<$0&$\sim$0&$\sim$0&$\sim$0&$>0$   &
$\sim$0&          \\
\end{tabular}
$^a$ T$_c$ = 21 K

$^b$ T$_c$ = 23 K, from Ref. 21
\end{table}

\end{document}